\documentstyle[twoside,fleqn,espcrc2]{article}

\newcommand{\AmS}{{\protect\the\textfont2
  A\kern-.1667em\lower.5ex\hbox{M}\kern-.125emS}}

\title{%
\begin{flushright} {\small HUPD-9416} \end{flushright}
        Construction of staples in lattice gauge theory
        on a parallel computer}

\author{S.Hioki\address{Department of Physics, Hiroshima University,
        Higashi-Hiroshima 724, Japan}%
        \thanks{To attend this conference, the author is financially
           supported by Electric Technology Research Foundation of Chugoku.}
        }

\begin{document}
\pagestyle{empty}

\begin{abstract}
        We propose a simple method to construct staples in lattice
gauge theory with Wilson action on a parallel computer.
        This method can be applicable to any dimensional system and
to any dimensional division without difficulty.
        Furthermore this requires rather small working area
to realize gauge simulation on a parallel computer.

\end{abstract}

\maketitle
\thispagestyle{empty}

\section{Introduction}

 Recently numerical simulations in lattice gauge theory become very important
to investigate non-perturbative physics as well as to evaluate interesting
quantities from first principle.
 In fact numerical studies grows with the growth of powerful computers like
supercomputers and parallel computers.
 There are many commercial parallel computers available now
as well as many dedicated QCD parallel computers.
 In particular many lattice QCD(Quantum Chromo Dynamics)
works have been performed on parallel computers\cite{Iwasaki94}.

 However for programmers, the variety of parallel computers prevents
the portability of their code.
 Moreover the parallel code usually needs more working area coming from
the overlapping with next processors than the code
which runs on a single processor.
 These "portability of code" and "reducing of working area" become
crucial to simulate a larger system and on a different platform.

 In this paper a simple method to overcome these problems in lattice QCD
Monte Carlo Simulations is proposed.
 The organization of this paper is as follows.
 In section 2, the notations and preparations which is necessary to
latter presentations is introduced.
 Actual construction of staples on a parallel computer is presented
in section 3.
 The summary is given in section 4.

\section{Notations and Preparations}

	Consider $d$-dimensional pure gauge theory defined on a lattice
with Wilson action.
	Let $U_{n,\mu}$ be a link variable of gauge field connecting
between site $n$ and $n + \hat{\mu}$,
the staple $X_{n,\mu}$ concerning the link $U_{n,\mu}$ is
\begin{eqnarray}
     X_{n,\mu}       &=&
     \sum_{\nu \neq \mu} ( X^{(+)}_{n,\mu \nu} + X^{(-)}_{n,\mu \nu} ), \\
     X^{(+)}_{n,\mu \nu} &=&
     U_{n,\nu} U_{n+\hat{\nu},\mu} U^{\dagger}_{n+\hat{\mu},\nu},  \\
     X^{(-)}_{n,\mu \nu} &=&
     U^{\dagger}_{n-\hat{\nu},\nu} U_{n-\hat{\nu},\mu}
       U_{n-\hat{\nu}+\hat{\mu},\nu} ,
\end{eqnarray}
making the Wilson action $S$ as
\begin{equation}
 S \propto \sum_{n,\mu} {\rm Re Tr} ( U_{n,\mu} X^{\dagger}_{n,\mu} ).
\end{equation}

	Let $N_{\mu}$ ($1 \le \mu \le d$) be a lattice size of
$\mu$ direction, then each site $x$ can be specified with its coordinate
$x=(x_1,x_2,...,x_d)$ ($0 \le x_{\mu} \le N_{\mu} - 1$),
and total volume is $\prod_{\mu=1}^d N_{\mu}$.
	Using these coordinates we can define a global address of site.
	One simple way to do this is to specify global address $j$
of site $x$ as
\begin{equation}
 j(x) = 1 + x_1 + \sum_{\mu=2}^d x_{\mu}  \prod_{k=1}^{\mu-1} N_k.
\end{equation}

	Then we introduce "normal order" of sites such that sites
are ordered with an increasing order of their global addresses.

	In order to realize even-odd (checkerboard) decomposition of sites
which is necessary for parallelization
we restrict on the case that each $N_{\mu}$ is even.
	We then call a site $x$ is even (odd) site
if $\sum_{\mu=1}^d x_{\mu} $ is even (odd).

	Next we put this lattice on a parallel computer which consists of
$\prod_{\mu=1}^d M_{\mu}$ ($1 \le M_{\mu} \le N_{\mu}$) processors, namely
we consider the case that each processor is responsible for
$\prod_{\mu=1}^d m_{\mu}$ ($m_{\mu} \equiv N_{\mu}/M_{\mu}$) sites.
	If $M_{\mu} \neq 1$,
$\mu$ direction diveded into $M_{\mu}$ pieces and put on multi processors.
	On the other hand if $M_{\mu} = 1$, $\mu$ direction is not divided
into processors but put on a single processor.
	In order to realize even-odd decomposition of sites also
in each processor, $\prod_{i=1}^d m_i$ is restricted to be even.

	From now on we restrict ourselves on a processor which is
denoted by ${\cal P}$,
but the story discussed below is universal and applicable to all processors
we think.
	For convenience we denote a neighboring processor
in $\pm \mu$ direction as ${\cal P} \pm \hat{\mu}$.

	Then we classify all sites in ${\cal P}$ into two groups;
one is even site group $G_e$ which consists of all even sites and the
other is odd site group $G_o$ which consists of all odd sites.
	Since both groups have $V$ ($\equiv \prod_{i=1}^d m_i / 2$) elements,
	we can number all elements of $G_e$ ($G_o$) in ${\cal P}$
from 1 to $V$ in normal order introduced above.
	In terms of this numbering
we can define a local address of sites in ${\cal P}$.

	Suppose a site n is in $G_e$ ($G_o$) and has a local address $l$.
	We refer the link variable $U_{n,\mu}$ as:
\begin{equation}
 U_{n,\mu} = \left\{
		 \begin{array}{@{\,}ll}
		   E_{l,\mu} & \mbox{( $n \in G_e$ )} \\
		   O_{l,\mu} & \mbox{( $n \in G_o$ )}
		 \end{array}
            \right. ,
\end{equation}
where $E_{l,\mu}$ ($O_{l,\mu}$) represents a link variable
attaching a even (odd) site whose address is $l$ and having a direction $\mu$.

	Next we introduce a list vector $v_e(l,\mu)$ ($v_o(l,\mu)$)
which points to a odd (even) site address located on $n + \hat{\mu}$
as a function of a even (odd) site address $l$.
	If $M_{\mu} = 1$, a site $n + \hat{\mu}$ is also
in ${\cal P}$.
	In this case $v_e(l,\mu)$ ($v_o(l,\mu)$) can be defined properly.
	If $M_{\mu} \neq 1$, however, there is a subgroup
$B^{(+)}_{e,\mu}$ ($B^{(+)}_{o,\mu}$) of $G_e$ ($G_o$) whose neighboring sites
in $+\mu$ direction are not in ${\cal P}$ but in ${\cal P} + \hat{\mu}$, i.e.
\begin{eqnarray}
     n \in B^{(+)}_{e,\mu} \cap G_e
                &\Rightarrow& n+\hat{\mu} \notin G_o \\
 \nonumber
     n \in {\bar B^{(+)}_{e,\mu}} \cap G_e
                &\Rightarrow& n+\hat{\mu} \in G_o \\
 \nonumber
     n \in B^{(+)}_{o,\mu} \cap G_o
                &\Rightarrow& n+\hat{\mu} \notin G_e \\
 \nonumber
     n \in {\bar B^{(+)}_{o,\mu}} \cap G_o
                &\Rightarrow& n+\hat{\mu} \in G_e \ .
\end{eqnarray}

	In this case $v_e(l,\mu)$ ($v_o(l,\mu)$) can not be defined
properly in ${\cal P}$ and we must extend the definition.
	We introduce a new group $N_{e,\mu}$ ($N_{o,\mu}$)
which consists of sites pointed by $n+\hat{\mu}$ where $ n \in B^{(+)}_{e,\mu}$
($n \in B^{(+)}_{o,\mu}$).
	It is noted that elements of $N_{e,\mu}$ ($N_{o,\mu}$) are not
in ${\cal P}$, but in ${\cal P} + \hat{\mu}$.
	Here we extend the site numbering in ${\cal P}$
to $N_{e,\mu}$ ($N_{o,\mu}$),
namely we number elements of $N_{e,\mu}$ ($N_{o,\mu}$) from
$V + 1$ to $V + V /m_{\mu}$ in normal order,
then we can define $v_e(l,\mu)$ ($v_o(l,\mu)$) which
points to a odd (even) site address located on $n + \hat{\mu}$ in
$N_{o,\mu}$ ($N_{e,\mu}$).

	Just like as $B^{(+)}_{e,\mu} (B^{(+)}_{o,\mu}$),
we can introduce a subgroup
$B^{(-)}_{e,\mu}$ ($B^{(-)}_{o,\mu}$) of $G_e$ ($G_o$) whose neighboring sites
in negative $\mu$ direction are not in
${\cal P}$ but in ${\cal P} - \hat{\mu}$.
	Since there are $V /m_{\mu}$ sites in
$B^{(-)}_{e,\mu}$ ($B^{(-)}_{o,\mu}$),
we can number these sites in $B^{(-)}_{e,\mu}$ ($B^{(-)}_{o,\mu}$) from
$i=1$ to $V /m_{\mu}$ in normal order.
	If a site n in $B^{(-)}_{e,\mu}$ ($B^{(-)}_{o,\mu}$)
has a local address $l$ and specified by a number $i$ discussed above,
we can introduce a new function $b_o(i,\mu) (b_e(i,\mu))$ defined as:
\begin{eqnarray}
  l & = & b_e(i,\mu) \ \ \ \ {\rm in} \ \ B^{(-)}_{e,\mu}, \ \ \ \
  1 \leq i \leq V /m_{\mu} \\
  \nonumber
  l & = & b_o(i,\mu) \ \ \ \ {\rm in} \ \ B^{(-)}_{o,\mu}, \ \ \ \
  1 \leq i \leq V /m_{\mu}.
\end{eqnarray}

\section{Construction of Staples}

	In this section, we construct staples.
For simplicity, we restrict on the case that $n \in G_e$.

\subsection{positive $\nu$ : $X^{(+)}_{n,\mu \nu}$}

	First we construct $X^{(+)}_{n,\mu \nu}$ in two steps.
We introduce temporary matrix T as:
\begin{equation}
 T = U_{n,\nu} U_{n+\hat{\nu},\mu}
\end{equation}
then we construct $X^{(+)}_{n,\mu \nu}$ as:
\begin{equation}
\label{eq:X+}
 X^{(+)}_{n,\mu} = T U^{\dagger}_{n+\hat{\mu},\nu}.
\end{equation}

	If $M_{\nu} \neq 1$ and in case of $n \in B^{(+)}_{e,\nu}$
we do not have $U_{n+\hat{\nu},\mu}$ in ${\cal P}$.
	Since the way to obtain information from other processors is
machine dependent, in this paper we use send/receive syntax to perform
inter-processor communications.
	We have to get $U_{n+\hat{\nu},\mu}$ from ${\cal P} + \hat{\nu}$,
in other words we have to send $U_{n+\hat{\nu},\mu}$ to ${\cal P} - \hat{\nu}$.
	So we first prepare the links that should be sent so that the links
are orderd sequentially in a dimension.
	Second we send links prepared to ${\cal P} - \hat{\nu}$.
	Third we receive links from ${\cal P} + \hat{\nu}$ and store them
to appropriate point of dimension.

\begin{center}
\begin{tabular}{lccc}
  PREPARE & $O_{\mu}(V+i)$ & $\leftarrow$ & $O_{\mu}(b_o(i,\nu))$ \\
  SEND    &  TO  & ${\cal P} - \hat{\nu}$ & $O_{\mu}(V+i)$   \\
 RECEIVE  &  FROM &  ${\cal P} + \hat{\nu}$ & $O_{\mu}(V+i)$
\end{tabular}
\end{center}

	For latter convenience, we introduce a function SETLINK that performs
above three operations. The syntax is:
\begin{equation}
 {\rm SETLINK}(O_{\mu},b_o,V,m_{\nu},\nu).
\end{equation}

	Now $T$ can be made by;
\begin{equation}
 T(l) = E_{\nu}(l) \ O_{\mu}(v_e(l,\nu)), \ \ \ 1 \leq l \leq V.
\end{equation}
	Obviously SETLINK is not necessary if $M_{\nu} = 1$.

	Next we prepare $U$s in eq.(\ref{eq:X+}).
If $M_{\mu} \neq 1$ first we perform SETLINK:
\begin{equation}
 {\rm SETLINK}(O_{\nu},b_o,V,m_{\mu},\mu).
\end{equation}
	Then we can construct $X^{(+)}$ as,
\begin{equation}
 X^{(+)}_{l,\mu \nu} = T(l) \ O^{\dagger}_{\nu}(v_e(l,\mu)),
 \ \ \ 1 \leq l \leq V.
\end{equation}

\subsection{negative $\nu$ : $X^{(-)}_{n,\mu \nu}$}

	To calculate $X^{(-)}_{n,\mu \nu}$ we use a technique.
	First construct $X^{(-)}_{n,\mu \nu}$ as $n - \hat{\nu}$
are the starting points.
	Next we move staples to $+\nu$ direction and obtain
$X^{(-)}_{n,\mu \nu}$.
	Just like as $X^{(+)}_{n,\mu \nu}$ case, we do in two steps.
	Since $n - \hat{\nu}$ are odd sites, temporal matrix T which
is the products of first two matrix of the right hand site of eq.(3)
can be construct as;
\begin{equation}
 T(l) = O^{\dagger}_{\nu}(l)\ O_{\mu}(l) , \ \ \ 1 \leq l \leq V.
\end{equation}

	If $M_{\mu} \neq 1$  do
\begin{equation}
 {\rm SETLINK}(E_{\nu},b_e,V,m_{\mu},\mu).
\end{equation}
	We then get $W(l)$ as;
\begin{equation}
 W(l) = T(l) \ E_{\nu}(v_o(l,\mu)) , \ \ \ 1 \leq l \leq V.
\end{equation}
which corresponds $X^{(-)}_{n,\mu \nu}$ but $n - \hat{\nu}$ are the
starting points instead of $n$s.
	Next we move $W(l)$ to $+\nu$ direction;
\begin{equation}
 T(v_o(l,\nu)) \leftarrow W(l) , \ \ \ 1 \leq l \leq V.
\end{equation}
	If $M_{\nu} \neq 1$ above $+\nu$ operation require inter-processor
communications. This can be done by sending $T(n+\nu)$
($n \in N_{e,\nu}$) to ${\cal P} + \hat{\nu}$
and by receiving them from to ${\cal P} - \hat{\nu}$
and storing them to $T(n)$ ($n \in B^{(-)}_{e,\nu}$).

\begin{center}
\begin{tabular}{lcccc}
  SEND    &  TO  & ${\cal P} + \hat{\nu}$ & $T(V+i)$
          \\
  RECEIVE &  FROM &  ${\cal P} - \hat{\nu}$
	  & $T(V+i)$
          \\
  STORE   & $T(b_e(i,\nu))$ & $\leftarrow$
          & $T(V+i)$
\end{tabular}
\end{center}
	This gives $X^{(-)}_{l,\mu \nu}$;
\begin{equation}
 X^{(-)}_{l,\mu \nu} = T(l), \ \ \ 1 \leq l \leq V.
\end{equation}

	Just like as SETLINK, it is useful to introduce a function
SLIDEMATRIX that performs above three operations.
	The syntax is:
\begin{equation}
 {\rm SLIDEMATRIX}(T,b_e,V,m_{\nu},\nu).
\end{equation}

\section{Summary}

 We have applied this method both on Fujitsu AP1000 at
Fujitsu Parallel Computing Research Facilities
and on Intel Paragon at Institute for Numerical Simulations
and Applied Mathematics in Hiroshima University.
We have checked that this method can be applicable to any dimensional system
and to any dimensional division of the original system without any difficulty.
Reduction of working area has been performed nicely compared with a original
program which uses overlapping with next processors.
 We hope that this method will be a good help
for beginners of parallel programming.

\section*{Acknowledgement}

 This work is financially supported by
Electric Technology Research Foundation of Chugoku.
 The author would like to thank the members of both Fujitsu Parallel Computing
Research Facilities and Supercomputer System Division of Intel Japan K.K.
for their technical support for parallel computing.
 He also thanks Dr. A.Nakamura for finding a bug of this program in the early
stage.

\end{document}